\documentclass[aps,twocolumn,noshowpaces,floatfix]{revtex4}
\usepackage{amsmath}
\usepackage{times}
\usepackage[dvips]{graphicx}
\usepackage[usenames]{color}

\usepackage{braket}

\begin{document}
\title{Finite-temperature time-dependent variation with multiple Davydov states}

\author{Lu Wang$^{1,2}$, Yuta Fujihashi$^{2}$,
Lipeng Chen$^{2}$,
Yang Zhao$^{2}$\footnote{Electronic address:~\url{YZhao@ntu.edu.sg}} }
\affiliation{$^1$Department of Physics, Zhejiang University, Hangzhou 310027, People's Republic of China\\
$^2$Division of Materials Science, Nanyang Technological University, Singapore 639798, Singapore}

\date{\today}

%%%%%%%%%%%%%%%%%%%%%%%%%%%%%%%%%%%%%%%%
\begin{abstract}
The Dirac-Frenkel time-dependent variational approach with Davydov {\it Ans\"atze} is a sophisticated, yet efficient technique to
obtain an accurate solution to many-body Schr\"odinger equations for energy and charge transfer dynamics in molecular aggregates
and light-harvesting complexes. We extend this variational approach to finite temperatures dynamics of the spin-boson model
by adopting a Monte Carlo importance sampling method.
In order to demonstrate the applicability of this approach, we compare real-time quantum dynamics of the spin-boson model calculated with that from numerically exact iterative quasiadiabatic propagator path integral (QUAPI) technique.
The comparison shows that our variational approach with the single Davydov {\it Ans\"atze} is in excellent agreement with the QUAPI method at high temperatures, while the two differ at low temperatures.
Accuracy in dynamics calculations employing a multitude of Davydov trial states is found to improve substantially over the single Davydov {\it Ansatz}, especially at low temperatures. At a moderate computational cost,
our variational approach with the multiple Davydov {\it Ansatz} is shown to provide accurate spin-boson dynamics over a wide range of temperatures and bath spectral densities.
\end{abstract}

\maketitle

%%%%%%%%%%%%%%%%%%%%%%%%%%%%%%%%%%%%%%%%
\section{Introduction}%%%%%%%%%%%%%%%%%%%%%%%%%
Responsible for electronic dephasing and energy relaxation, interplay between electronic and nuclear degrees of freedom (DOFs) is a fundamental aspect of dynamical processes in condensed phases such as molecular aggregates and light-harvesting complexes\cite{nitzan2006chemical,may2011charge}.
An accurate description of quantum dissipative dynamics in the condensed phases remains a challenging problem.
One of main schemes for treating quantum dissipation is the reduced density matrix approach, with a focus on truncated system dynamics in the presence of a macroscopic thermal bath.
The second-order cumulant time-nonlocal quantum master equation approach\cite{tanimura1989time,ishizaki2005quantum,ishizaki2009unified,fujihashi2015impact,chen2015dynamics} and path integral methods such as iterative quasiadiabatic propagator path integral (QUAPI) technique\cite{makri1995tensor1,makri1995tensor2,nalbach2011iterative} are examples of numerically exact methods for propagating the reduced density matrix.
These nonperturbative and non-Markovian approaches allow for exploration of a full range of system-bath coupling and electronic coupling strengths\cite{nalbach2011iterative,ishizaki2009unified} although it becomes extremely difficult to predict quantum dissipative dynamics at very low temperatures.\cite{tang2015extended,chen2016inchworm}
Recent progress in ultrafast time resolved spectroscopy has stimulated methodological developments, and a large number of efficient, approximate reduced density matrix approaches are available.\cite{cheng2009dynamics,ishizaki2010quantum,huelga2013vibrations,chenu2015coherence,chen2015optimal,lee2016semiclassical}

In the aforementioned reduced density matrix approaches, coupled quantum dynamics of electronic and bosonic degrees of freedom (DOFs) is obtained explicitly only for electronic systems, whereas the bath DOFs are traced out in the reduced density matrix.\cite{may2011charge}
%{\bf Because of this treatment, all explicit information of the bath dynamics is lost, and the interplay between the system and the bath is only reflected in system observables such as optical spectra and electronic populations. 
%Recent ultrafast nonlinear spectroscopic techniques, such as femtosecond stimulated Raman spectroscopy\cite{kukura2007femtosecond} and two-dimensional electronic-vibrational spectroscopy,\cite{oliver2014correlating} allow direct monitoring of the temporal evolution of nuclei in the ground or excited electronic states. 
%Hence it becomes increasingly important to address explicit boson dynamics of a many-body boson-electronic system in order to fully interpret nonlinear spectroscopic signals.} 
Alternatively, the wave-function approach such as the multiconfigurational time-dependent Hartree (MCTDH) approach\cite{meyer1990multi,matzkies1999accurate} and its multilayer variant (ML-MCTDH)\cite{wang2001systematic,wang2003multilayer,manthe2008multilayer,wang2006quantum} can describe time propagation of the wave function of all DOFs explicitly, with equations of motion determined by the Dirac-Frenkel time-dependent variational principle. 
These approaches are shown to be powerful tools for obtaining numerically exact quantum dynamics at very low temperatures,\cite{wang2003multilayer} and for including finite temperature effects with the help of statistical sampling of the bath initial conditions.\cite{wang2006quantum}

One of the established approximate methods for describing time evolution of many-body wave functions is to employ the Dirac-Frenkel time-dependent variational approach with the Davydov {\it Ans\"atze}, \cite{vskrinjar1988classical,forner1996accuracy,sun2010dynamics} which consist of sums of direct product states of localized electronic state and coherent states of the bath modes as proposed by Davydov in 1970s to describe a soliton motion in molecular chains.\cite{davydov1979solitons}
The time-dependent variational approach with these {\it Ans\"atze} has been widely used for describing excitation dynamics and nonlinear optical spectra of Holstein polaron and molecular aggregates.\cite{huynh2013polaron,chorovsajev2014dynamics,chen2015theory}
Recently it has been shown that superpositions of the Davydov {\it Ans\"atze}, also known as the multiple Davydov {\it Ans\"atze}, provide significant improvements in the accuracy of the time-dependent variation.\cite{zhou2014ground,zhou2015symmetry,zhou2015polaron,wang2016variational,huang2016polaron,zhou2016fast,deng2016dynamics,fujihashi2016effect}
Increasing the multiplicity of these {\it Ans\"atze} leads to numerically exact dynamics in the Holstein model and the sub-Ohmic spin boson model.\cite{wang2016variational,huang2016polaron,zhou2016fast,deng2016dynamics}
However, the usual variational approach with the Davydov {\it Ans\"atze} has been restricted to zero temperature case because of the wave function formalism.

In this study, we remove the restriction of zero temperature of the variational approach with the Davydov {\it Ans\"atze} in order to include the effect of finite temperature.
For this purpose, we adopt a Monte Carlo like method, inspired by the ML-MCTDH approach.\cite{wang2006quantum}
The reliability of the extended variational approach with the Davydov {\it Ansatz} at finite temperature is identified through comparison with benchmark results obtained from the QUAPI approach.
The rest of the paper is organized as follows.
In Sec.~II, we introduce the spin-boson Hamiltonian, Davydov {\it Ans\"atze}, and Monte Carlo importance sampling method.
In Sec.~III, we present a comparison of population dynamics between our theoretical approach and the numerically accurate QUAPI results at finite temperatures.
Finally, Sec.~IV is devoted to concluding remarks.

\setcounter{equation}{0}

\section{Model\label{sec:Section-II}}%%%%%%%%%%%%%%%%%%%%%%%%
\label{sec:METHODOLOGY}
\subsection{Hamiltonian}%%%%%%%%%%

In this study, we consider a two-level system coupled to a single dissipative bath composed of harmonic oscillators as described by the spin-boson model.\cite{leggett1987dynamics}
The spin-boson model is a paradigm for studying a variety of physical and chemical phenomena such as electron transfer and exciton dynamics in condensed phase systems.\cite{nitzan2006chemical,weiss2012quantum}
Thus it is suitable for a benchmarking system for investigating the validity of the newly developed quantum dynamical approach.
The Hamiltonian of the spin-boson model is written as (we set $\hbar =1$)
\begin{align}
\hat{H}= \hat{H}_{\rm S} + \hat{H}_{\rm SB} + \hat{H}_{\rm B},
\end{align}
where
\begin{align}
\hat{H}_{\rm S}=  \frac{\epsilon}{2} \sigma_z - \frac{\Delta}{2} \sigma_x ,
\end{align}
\begin{align}
\hat{H}_{\rm SB}=
\frac{\sigma_z }{2} \sum_l \lambda_l (b_l^\dagger + b_l),
\end{align}
\begin{align}
\hat{H}_{\rm B}= \sum_l \omega_l b_l^\dagger b_l.
\end{align}
Here, $\sigma_{i}$ ($i=x,z$) are Pauli operators defined as $\sigma_x = | 1\rangle \langle 2| + | 2\rangle \langle 1 | $ and $\sigma_z = | 1\rangle \langle 1| - | 2\rangle \langle 2 | $ with $ | 1\rangle$ and $| 2\rangle$ representing two electronic localized states.
$\epsilon$ and $\Delta$ is the energy bias and the coupling constant between two electronic states, respectively.
$b_l^\dagger$ ($b_l$) is creation (annihilation) operator for the bosonic bath mode of frequency $\omega_l$, and $\lambda_l$ is the strength of the coupling between the system and the $l$th mode.

Owing to Wick's theorem,\cite{wick1950evaluation,fetter2003quantum} all the fluctuations and the dissipation process of $\hat{u}(t)\equiv   e^{i \hat{H}_{\rm B}t} \sum_l  \lambda_l (b_l^\dagger + b_l) e^{-i \hat{H}_{\rm B}t}$ can be specified by a two-body correlation function:\cite{kubo1985statistical}
\begin{align}
C(t)=&\langle \hat{u}(t) \hat{u}(0) \rangle_{\rm B} \notag \\
	=& \int_0^\infty d\omega J(\omega) [\coth({\beta \omega/2}) \cos \omega t - i \sin \omega t],
\end{align}
where $\langle \dots \rangle_{\rm B}$ denotes averaging over $\hat{\rho}_{\rm B}^{\rm eq}=e^{-\beta \hat{H}_{\rm B} }/ {\rm Tr_B} \{ e^{-\beta \hat{H}_{\rm B}} \}$ with $\beta=k_{\rm B}T$ being the inverse temperature.
The bath spectral density is defined in terms of the coupling strength $\lambda_l$ as\cite{may2011charge}
\begin{align}
J(\omega)=\sum_l \lambda_l^2  \delta(\omega -\omega_l).
\end{align}
Several forms of $J(\omega)$ are employed in the literature based on model assumptions.
In this study, we adopt the following spectral density form
\begin{align}
J(\omega)=2\alpha \omega_c^{1-s} \omega^s e^{-\omega/ \omega_c}, \label{eq6}
\end{align}
where $\alpha$ represents the strength of the coupling, and $\omega_c$ provides a phenomenological frequency cutoff.
Those spectral densities with $s < 1$, $s = 1$, and $s > 1$ are referred to as sub-Ohmic, Ohmic, and super-Ohmic, respectively.\cite{leggett1987dynamics}

To obtain numerical solutions to the equations of motion for the variational parameters, the continuum spectral density needs to be discretized.
The conventional logarithmic discretization method has been employed to investigate the sub-Ohmic spin-boson model at zero temperature,\cite{bulla2003numerical,zhou2014ground,zhou2015symmetry,wang2016variational} but may not be appropriate for finite temperatures.
The logarithmic discretization method samples more in the low frequency domain, and can easily characterize the low frequency bath modes at zero temperature.
On the other hand, the high frequency bath modes should also be of importance at finite temperatures due to their initial excitations  according to the Bose statistics.
For this reason, we follow the spectral density discretization procedure in the ML-MCTDH approach.\cite{wang2001systematic,wang2003multilayer}
Following Ref.~\onlinecite{wang2003multilayer}, we introduce a density of frequencies $\Xi(\omega)$ defined on $[0,\omega_{\rm max}]$, where $\omega_{\rm max}$ is the upper bound of the frequency, and discretize the continuum of frequencies as
\begin{equation}
  \label{eq:dis_freq}
  \int_{0}^{\omega_{l}} d\omega\ \Xi(\omega) = l,\ l=1,\, 2,\, \dots,\, N_{b},
\end{equation}
where $N_{b}$ is the number of discrete bath modes and $\omega_{N_{b}} = \omega_{\max}$.
The parameter $\lambda_{l}$ for each $\omega_{l}$ is then given by $ \lambda_{l} = \sqrt{J(\omega_{l}) / \Xi(\omega_{l})}$.
The precise functional form of $\Xi(\omega)$ does not affect the final outcome if sufficient bath modes are included.
For efficient numerical calculations, however, the form of $\Xi(\omega)$ can be chosen as
\begin{equation}
  \label{eq:density}
  \Xi(\omega) = \frac{1} {\Gamma_{\omega_{\max}}} \frac{J(\omega)} {\omega},
\end{equation}
where
\begin{equation}
  \label{eq:den_norm}
  \Gamma_{\omega_{\max}} = \frac{1} {N_{b}} \int_{0}^{\omega_{\max}} d\omega\ \frac{J(\omega)} {\omega}.
\end{equation}
Here, $ \Gamma_{\omega_{\max}}$ is the factor guaranteeing $\int_{0}^{\omega_{\max}} d\omega\ \Xi(\omega) = N_{b}$.
For the Ohmic bath ($s=1$), $\omega_{l}$ is expressed as $\omega_{l} = -\omega_{c} \ln \left( 1- l \Gamma_{\omega_{\max}} / \omega_c\right)$.
For the sub-Ohmic bath ($s<1$), $\omega_{l}$ can only be given implicitly through Eq.~(\ref{eq:dis_freq}).

It is noted that an arbitrary spectral density can be employed in our variational approach.
However, handling the spectral density with a complex structure may require a large number of discretized bath modes or a sophisticated discretization procedure.
Thus, for simplicity we restrict our numerical calculations to the functional form of Eq.~(\ref{eq6}) since the goal of our study is to demonstrate the validity of our finite temperature approach.

\subsection{Time dependent variational approach with multiple Davydov trial states}%%%%%%%%

In general, the time evolution of the wave function $| \psi (t) \rangle$ for the total Hamiltonian, $\hat{H}$, is described with the schr{\"o}dinger equation,
\begin{align}
i  \frac{\partial }{\partial t} | \psi (t) \rangle= \hat{H}  | \psi (t) \rangle .
\end{align}
In this work, we employ two trial wave functions, namely, the multiple $\mathrm{D}_{1}$ and $\mathrm{D}_{2}$ {\it Ansatz}, to solve the above schr{\"o}dinger equation. Both are known as the multiple Davydov {\it Ans\"atze}.
An extension to the single $\mathrm{D}_{1}$ {\it Ansatz}, the multiple $\mathrm{D}_{1}$ {\it Ansatz} has been employed to investigate both static and dynamical properties of many-body quantum systems such as the spin-boson model and the Holstein molecular crystal model.\cite{zhou2015symmetry,wang2016variational,zhou2016fast}
The time-dependent version of the multi-$\mathrm{D}_{1}$ {\it Ansatz} can be written as
\begin{eqnarray}
  \Ket{\mathrm{D}_1^M(t)}&=&\ket{1}\sum_{n=1}^{M}A_{n}(t)\exp\left(\sum_{l}f_{nl}(t)b_{l}^{\dagger}-\mathrm{{H.c.}}\right)\ket{0}_{\rm B}
  \nonumber\\
  &&+\ket{2}\sum_{n=1}^{M}B_{n}(t)\exp\left(\sum_{l} g_{nl}(t)b_{l}^{\dagger}-\mathrm{{H.c.}}\right)\ket{0}_{\rm B}, \nonumber\\
  \label{D1fun}
\end{eqnarray}
where $\ket{0}_{\rm B}$ is the vacuum state of the boson bath.
The variational parameters $A_{n}(t)$ and $B_{n}(t)$ are the amplitudes in states $\ket{1}$ and $\ket{2}$, respectively.
$f_{nl}(t)$ and $g_{nl}(t)$ represent bath-mode displacements with $n$ and $l$ denoting the $n$th coherent state and the $l$th  bath mode, respectively.
$M$ is the multiplicity, which denotes the number of single $\mathrm{D}_{1}$ states in Eq.~(\ref{D1fun}).
For $M=1$, the multi-$\mathrm{D}_{1}$ {\it Ansatz} reduces to the standard Davydov $\mathrm{D}_{1}$ {\it Ansatz}.
Similarly, the multi-$\mathrm{D}_{2}$ {\it Ansatz} is given by\cite{zhou2015polaron,huang2016polaron}
\begin{eqnarray}
  \Ket{\mathrm{D}_2^M(t)}&=&\ket{1}\sum_{n=1}^{M}A_{n}(t)\exp\left(\sum_{l}f_{nl}(t)b_{l}^{\dagger}-\mathrm{{H.c.}}\right)\ket{0}_{\rm B}
  \nonumber\\
  &&+\ket{2}\sum_{n=1}^{M}B_{n}(t)\exp\left(\sum_{l} f_{nl}(t)b_{l}^{\dagger}-\mathrm{{H.c.}}\right)\ket{0}_{\rm B}, \nonumber\\
\end{eqnarray}
where $f_{nl}(t)$ are the displacements of $l$th bath mode on any electronic states.
Thus, the multi-$\mathrm{D}_{2}$ {\it Ansatz} is a simplified version of the multi-$\mathrm{D}_{1}$ {\it Ansatz}, since the bath displacements of the multiple $\mathrm{D}_{1}$ ($\mathrm{D}_{2}$) trial state is site-dependent (site-independent).
For $M=1$, the multi-$\mathrm{D}_{2}$ {\it Ansatz} reduces to the single $\mathrm{D}_{2}$ {\it Ansatz}.

The time-dependent variational parameters, $A_{n}(t)$, $B_{n}(t)$, $f_{nl}(t)$ and $g_{nl}(t)$ are determined by adopting the Lagrangian formalism of the Dirac-Frenkel time-dependent variational principle.
The Lagrangian associated with the trial state $| {\rm D}_1^M(t) \rangle$ is given by
\begin{equation}
  \label{Lag}
  L=\frac{i}{2}  \langle \mathrm{D}_1^M(t) |\frac{\overset\leftrightarrow\partial}{\partial t}|\mathrm{D}_1^M(t) \rangle
  -  \langle \mathrm{D}_1^M(t)|\hat{H}|\mathrm{D}_1^M(t) \rangle,
\end{equation}
where the operator $\overset\leftrightarrow\partial / \partial t$ denotes
$\overset\rightarrow \partial/\partial t - \overset\leftarrow \partial/\partial t $.
The Dirac-Frenkel time-dependent variational principle yields the equations of motion for the variational
parameters,
\begin{eqnarray}
&&  \frac{d}{dt} \left( \frac{\partial L} {\partial \dot{u}_{n}^{\ast}} \right) - \frac{\partial L}{\partial u_{n}^{\ast}} = 0,
\label{dirac_frenkel}
\end{eqnarray}
where $u_{n}$ are the variational parameters, i.e., $A_{n}(t)$, $B_{n}(t)$, $f_{nl}(t)$ and $g_{nl}(t)$, and $u_{n}^{\ast}$ is the complex conjugate of $u_{n}$.
Similarly, time evolution for the multi-$\mathrm{D}_2$ {\it Ansatz}, can be derived using the Dirac-Frenkel time-dependent variational principle.
Detailed derivations of the equations of motion for the variational parameters of multiple Davydov {\it Ans\"atze} can be found in Ref.~\onlinecite{wang2016variational}.

The expectation value of an observable can be expressed as $\langle \hat{O}(t) \rangle = \mathrm{Tr} \{ \hat{O} \hat{\rho}_{\rm tot}(t)  \} $, where $\hat{\rho}_{\rm tot}(t)=e^{-i \hat{H}t}\hat{\rho}_{\rm tot}(0) e^{i \hat{H}t}$ is the total density operator, and $\hat{O}$ is a time independent operator.
Using the multi-$\mathrm{D}_1$ or $\mathrm{D}_2$ {\it Ansatz}, the expectation value of the observable of interest at zero temperature can be calculated as
\begin{align}
  \langle \hat{O}(t) \rangle  =
   \langle \mathrm{D}_{1,2}^M (t)  | \hat{O} | \mathrm{D}_{1,2}^M (t) \rangle . \label{eq5}
\end{align}
In the next subsection, this expression is extended to the finite temperatures.

\subsection{Observables at finite temperature}%%%%%%%%%%%%

The conventional time-dependent variation described above is only applicable at zero temperature.
In this subsection, we extend this variational approach to finite temperatures by adopting a Monte Carlo like method, similar to what is used in the ML-MCTDH approach.\cite{wang2003multilayer,wang2006quantum}
The initial density matrix for the entire system is assumed to have a factorized form, {\it i.e.}
$\hat{\rho}_{\rm tot}(0)=\hat{\rho}(0) \hat{\rho}_{\rm B}^{\rm eq}$, where $\hat{\rho}(0)=| 1 \rangle \langle 1 |$.
The extension to more general initial conditions with superposition of $| 2 \rangle $ and $| 1 \rangle $ is straightforward, which is important for modeling nonlinear spectroscopy.

The expectation value of an observable $\hat{O}(t)$ at finite temperatures can be expressed as
\begin{align}
\langle \hat{O}(t) \rangle = \mathrm{Tr} \{ \hat{O} e^{-i\hat{H} t} \hat{\rho}_{\rm B}^{\rm eq}| 1 \rangle \langle 1 |  e^{i\hat{H} t} \} .
\end{align}
In principle, the observables at $t$ can be calculated in any representations.
We employ the coherent state representation to calculate the observable as
\begin{align}
  \langle \hat{O}(t) \rangle  =  \pi^{-N_{b}} \int d^{2}  \mbox{\boldmath $\alpha$}
    \langle  \mbox{\boldmath $\alpha$} |   \langle 1 |  \hat{\rho}_{\rm B}^{\rm eq}
  e^{i \hat{H}t} \hat{O} e^{-i \hat{H}t}  | 1 \rangle | \mbox{\boldmath $\alpha$} \rangle, \label{eq1}
\end{align}
where $ | \mbox{\boldmath $\alpha$} \rangle $ denotes a direct product of coherent states $(\alpha_{1}$,~$\alpha_{2}$,~$\alpha_{3}$,~$\cdots$,~$\alpha_{N_{b}})$ for the $N_{b}$ discrete bath modes, and is expressed as $ | \mbox{\boldmath $\alpha$} \rangle =\exp (\sum_{l} \alpha_{l} b_{l}^{\dagger}-\mathrm{{H.c.}} )\ket{0}_{\rm B}$.
Each $\alpha_i$ runs over all of the feasible coherent states.
The element of area $d^{2} \alpha_{i}$ on the complex plane of $\alpha_{i}$ denotes $d\mathrm{Re}(\alpha_{i}) \cdot d\mathrm{Im}(\alpha_{i})$, which $\mathrm{Re}(\alpha_{i})$ and $\mathrm{Im}(\alpha_{i})$ are
the real and imaginary part of $\alpha_{i}$, respectively.  The equilibrium density matrix of
the bath at a finite temperature is a diagonal matrix, and can be expressed as\cite{glauber1963coherent,sudarshan1963equivalence}
\begin{align}
  \hat{\rho}_{\rm B}^{\rm eq}= \int d^{2}\mbox{\boldmath $\alpha$} \, p(\mbox{\boldmath $\alpha$} ;\, \beta)
  | \mbox{\boldmath $\alpha$} \rangle \langle \mbox{\boldmath $\alpha$}|, \label{eq2}
\end{align}
where $p(\mbox{\boldmath $\alpha$};\, \beta)$ represents the diagonal elements of the density matrix in the
coherent state representation and can be expressed as\cite{hillery1984distribution}
\begin{equation}
  p(\mbox{\boldmath $\alpha$};\, \beta) = \prod_{l}^{N_b} \left[ \frac{e^{\beta \omega_{l}} -1} {\pi} \exp
    \left( -|\alpha_{l}|^{2} (e^{\beta \omega_{l}} -1) \right) \right]. \label{eq3}
\end{equation}
As shown in Eq.~(\ref{eq3}), $p(\mbox{\boldmath $\alpha$};\, \beta)$ is a positive defined function of $\mbox{\boldmath $\alpha$}$ and can be seen as a probability
density.
Substituting Eq.~(\ref{eq2}) into Eq.~(\ref{eq1}), the observables $\braket{O(t)}$ at finite temperature can be obtained by the average according to the probability density $ p(\mbox{\boldmath $\alpha$};\, \beta)$ as
\begin{align}
  \langle \hat{O}(t) \rangle  =&  \int d^{2}  \mbox{\boldmath $\alpha$}     p(\mbox{\boldmath $\alpha$};\, \beta)
    \langle  \mbox{\boldmath $\alpha$} |     \langle 1 |
  e^{i \hat{H}t} \hat{O} e^{-i \hat{H}t} | 1 \rangle  | \mbox{\boldmath $\alpha$} \rangle \notag \\
  =& \int d^{2}  \mbox{\boldmath $\alpha$}     p(\mbox{\boldmath $\alpha$};\, \beta)
   \langle \mathrm{D}_{1,2}^M (t;\, \mbox{\boldmath $\alpha$})  | \hat{O} | \mathrm{D}_{1,2}^M (t;\,\mbox{\boldmath $\alpha$}) \rangle . \label{eq4}
\end{align}
For the second equality, we have use the multi-$\mathrm{D}_{1}$ or $\mathrm{D}_{2}$ {\it Ansatz}, $\ket{\mathrm{D}_{1,2}^M(t;\, \mbox{\boldmath $\alpha$})} = e^{-i \hat{H}t} \ket{1} \ket{\mbox{\boldmath $\alpha$}} $, and $\ket{\mathrm{D}_{1,2}^M(0;\, \mbox{\boldmath $\alpha$})}$ denotes a trial state with initial bath displacements of $\mbox{\boldmath $\alpha$}$ at $t=0$.
For the case of the multi-$\mathrm{D}_{1}$ {\it Ansatz}, initial condition parameters are $A_{1}(0)=1$, $B_{1}(0)=0$, $A_{n}(0)=B_{n}(0)=0$ for $n\neq 1$ and $f_{nl}(0) = g_{nl}(0) = \alpha_{l}$ for all $n$ and  $l$.
Likewise, initial parameters of the multi-$\mathrm{D}_{2}$ {\it Ansatz} are $A_{1}(0)=1$, $B_{1}(0)=0$, $A_{n}(0)=B_{n}(0)=0$ for $n\neq 1$ and $f_{nl}(0) = \alpha_{l}$ for all $n$ and  $l$.

The expectation value of the observable at finite temperature can numerically be calculated by the technique of Monte Carlo importance sampling as
\begin{align}
  \langle \hat{O}(t) \rangle  =\frac{1}{N_s} \sum_ {i}^{N_s}
   \langle \mathrm{D}_{1,2}^M (t;\, \mbox{\boldmath $\alpha$}_i)  | \hat{O} | \mathrm{D}_{1,2}^M (t;\,\mbox{\boldmath $\alpha$}_i) \rangle ,
   \label{eq8}
\end{align}
where $N_s$ is the sampling number.
The configuration $\mbox{\boldmath $\alpha$}_i$ for the bath is numerically generated according to $p(\mbox{\boldmath $\alpha$};\, \beta)$ by importance sampling, where $p(\mbox{\boldmath $\alpha$};\, \beta)$ is the Boltzmann distribution used as the weighting function in the importance sampling procedure.
Letting $2\sigma_l^2=1/(e^{\beta \omega_l} -1)$ and $\alpha_l= x_l+ip_l$, $ p(\mbox{\boldmath $\alpha$};\, \beta)$ in Eq.~(\ref{eq3}) can be partitioned into two independent Gaussian distribution as
\begin{align}
p(\mbox{\boldmath $\alpha$};\, \beta)= \prod_{l}^{N_b}
 \frac{1}{\sqrt{2\pi}\sigma_l} e^{-\frac{x_l^2}{2\sigma_l^2}}
\frac{1}{\sqrt{2\pi}\sigma_l} e^{-\frac{p_l^2}{2\sigma_l^2}} , \label{eq7}
\end{align}
where $\sigma_l$ can be taken as the variance of the Gaussian distribution.
To avoid singularity, the initial displacements in the trial states is determined by setting $f_{nl}(0)=g_{nl}(0)=\alpha_l + \epsilon_0$, where noise $\epsilon_0$ satisfying the uniform distribution $[-10^{-2},10^{-2}]$ is added to the variational parameters of the initial states.
From the definition of $\sigma_l$, a lower temperature or the higher frequency $\omega_l$ gives a smaller $\sigma_l$.
The zero temperature case corresponds to every bath mode being in the ground state initially, and it is equivalent to a coherent state with displacement parameter, $\alpha_l=0$ for all $l$.
In this case, the observable expression of Eq.~(\ref{eq4}) or Eq.~(\ref{eq8}) reduces to Eq.~(\ref{eq5}).

\section{Results and discussion\label{sec:Section-III}}%%%%%%%%%%%%%%%%%%%%%%%%

\setcounter{equation}{0}

\begin{figure}%%%%%%%%%%
	 \includegraphics{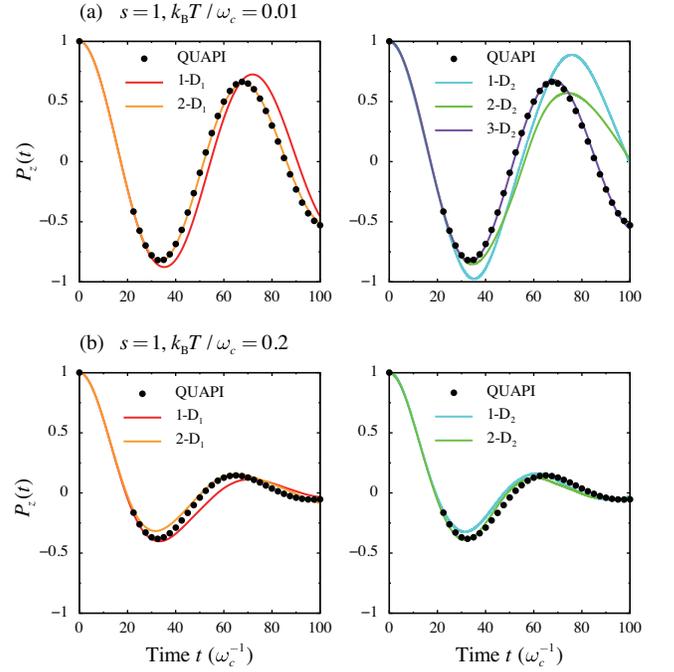}	

	\caption{
	Time evolution of the population difference $P_z(t)$ in Ohmic case ($s=1$).
	The factorized bath initial condition is employed.
	The results calculated by the multi-${\rm D}_1$ {\it Ans\"atze} (left panels, solid lines in red and orange) and the multi-${\rm D}_2$ {\it Ans\"atze} (right panels, solid lines in blue, green and purple) are plotted for (a) low temperature ($k_{\rm B}T/\omega_c=0.01$) and (b) high temperature case ($k_{\rm B}T/\omega_c=0.2$).
	The black circles indicate the QUAPI results with $\Delta t=0.125$ and $k_{\rm max}=7$.
	The other parameters are fixed to be $\epsilon=0$, $\Delta=0.1\omega_c$ and $\alpha=0.05$.}
	\label{fig1}
\end{figure}%%%%%%%%%%%%

In this section, we compare results from our variational approach with those of the QUAPI method\cite{makri1995tensor1,makri1995tensor2,nalbach2011iterative} in order to demonstrate the applicability of our approach to unraveling many-body dynamics.
In the spin-boson model, a principal observable of interest is the population difference between the electronic states, $P_z(t)$, calculated as
\begin{align}
P_z(t)=\langle \sigma_z (t)\rangle=
\frac{1}{N_s} \sum_ {i}^{N_s}
   \langle \mathrm{D}_{1,2}^M (t;\, \mbox{\boldmath $\alpha$}_i)  | \sigma_z | \mathrm{D}_{1,2}^M (t;\,\mbox{\boldmath $\alpha$}_i) \rangle .
\end{align}
All units in the numerical calculations are scaled by the cutoff frequency $\omega_c$.
In this section, for simplicity, we choose zero bias as a large energy bias requires a large multiplicity.\cite{borrelli2016quantum}
The electronic coupling constant and the coupling strength of spectral density are set as $\Delta=0.1\omega_c$ and $\alpha=0.05$, respectively.
The number of the discrete bath modes is fixed at $N_b=250$.
{\bf In all numerical calculations of this study}, statical averages are taken over a maximum of 400 realizations, {\bf which is the number sufficient for convergence at $k_{\rm B}T=0.2\omega_c$.
For lower temperatures the observables converge with less realizations.}

Here we briefly mention the numerical convergence of the QUAPI procedure.
The basic idea here is to consider a Trotter splitting of the short-time propagator for the total Hamiltonian into one part depending on the system Hamiltonian and another involving the bath and the system-bath coupling term.\cite{makri1995tensor1,makri1995tensor2,nalbach2011iterative}
The short-time density matrix propagator describes time evolution over a time slice $\Delta t$.
This splitting is by construction exact in the limit $\Delta t \rightarrow 0$.
With finite time slice $\Delta t$ in practical calculations, the numerical error can be eliminated by choosing sufficiently small $\Delta t$ to achieve convergence.
On the other hand, the bath DOFs generate bath correlations in the Feynman-Vernon influence functional.
For any finite temperature, these correlations decay with a dephasing time constant of $\omega_c^{-1}$, thus a memory time window, $\tau_{\rm mem}=k_{\rm max} \Delta t$, should be defined to handle the bath correlation truncation beyond a certain time span.
Neglected beyond $\tau_{\rm mem}$, all correlations are included exactly within a finite memory time of $\tau_{\rm mem}$.
To reach convergence, the memory time window should be enlarged by increasing $k_{\rm max}$ such that all memory effects are taken into account up to a desired accuracy.
The QUAPI procedure fails to converge when the memory time is too long.
The accuracy of the truncation of $k_{\rm max}$ in Figs.~\ref{fig1} and \ref{fig2} were checked to make sure that the numerical results are converged.
In this work, the time slice and the number of memory parameter are set to be $\Delta t=0.125$ and $k_{\rm max}=7$, respectively.
The convergence failure of the QUAPI in deep sub-Ohmic regime is discussed in Subsection~IIIB.

\subsection{The Ohmic Regime}%%%%%%%%%

Figures~\ref{fig1}(a) and \ref{fig1}(b) display time evolution of $P_z(t)$ calculated by the multi-$\rm D_1$ {\it Ans\"atze}, the multi-$\rm D_2$ {\it Ans\"atze} and the QUAPI approach in the Ohmic case ($s=1$) at two temperatures ($k_{\rm B}T/\omega_c=0.01$ and $k_{\rm B}T/\omega_c=0.2$), respectively.
There is perfect agreement between the ${\rm D}_2^{M=3}$ {\it Ansatz} and the QUAPI approach at the low temperature $k_{\rm B}T/\omega_c=0.01$, as shown in the right panel of Fig.~\ref{fig1}(a).
On the other hand, the agreement between the ${\rm D}_2^{M=1}$ {\it Ansatz} and the QUAPI approach is unsatisfactory.
As discussed in Ref.~\onlinecite{wang2016variational}, this is because time evolution of the phonon wave functions show the plane wave like behavior in the weak coupling regime and is difficult to be described by superposition of coherent states.
Thus, more phonon coherent states are needed to capture the accurate dynamics.
The multi-${\rm D}_1$ {\it Ansatz} provides accurate population dynamics by $M=2$, and the agreement between the multi-${\rm D}_1$ {\it Ansatz} and the QUAPI approach is good even at $M=1$.
This clearly demonstrates that the multi-${\rm D}_1$ {\it Ans\"atze} is more flexible than the multi-${\rm D}_2$ {\it Ans\"atze}.

In the left panel of Fig.~\ref{fig1}(b), dynamics obtained by both the ${\rm D}_1^{M=2}$ {\it Ansatz} and the QUAPI approach agree with each other at high temperature ($k_{\rm B}T/\omega_c=0.2$) despite small difference between the ${\rm D}_1^{M=2}$ {\it Ansatz} and QUAPI populations.
The result by the ${\rm D}_1^{M=1}$ {\it Ansatz} also appears to be similar to that by the $M=2$ case.
The small difference between the multi-${\rm D}_1$ {\it Ans\"atze} and the QUAPI approach may be due to insufficient number of bath modes.
From Eq.~(\ref{eq7}) and the definition of the variance of the Gaussian distribution $\sigma_l$,  the value of $\sigma_l$ becomes large at high temperature.
The increase of $\sigma_l$ leads to large values of $f_{nl}(0)$ and $g_{nl}(0)$ even for high frequency modes which can be ignored at low temperatures.
Therefore, the number $N_b$ of bath modes required as well as the sampling number become large due to large value of $\sigma_l$ at high temperatures.
The accuracy of the ${\rm D}_2$ {\it Ansatz} with $M=1$ at high temperature is improved significantly unlike in the low temperature regime, and are similar to that of the ${\rm D}_1^{M=2}$ {\it Ansatz}.
These results indicate that the superposition of the Davydov trial states is more fragile against thermal fluctuations at higher temperatures, and thus the trial states with small multiplicity are sufficient for description of dynamics in this regime.
The increased computational cost due to additional bath modes and extended sampling is offset by the reduced {\it Ansatz} multiplicity, and thus our variational approach with importance sampling remains efficient even at high temperatures.

\subsection{The Sub-Ohmic Regime}%%%%%%%%%

\begin{figure}%%%%%%%%%%

 	 \includegraphics{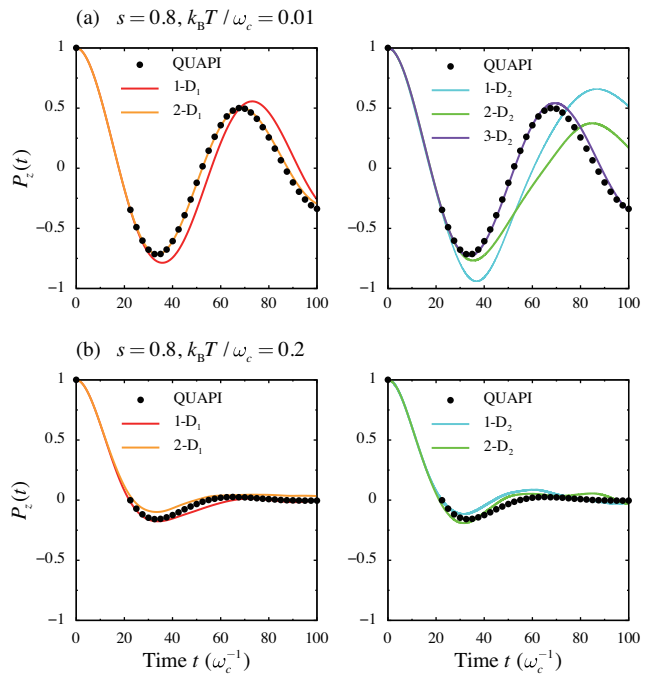}		

	\caption{
	Time evolution of the population difference $P_z(t)$ with spectral density exponent $s=0.8$.
	The factorized bath initial condition is employed.
	The results calculated by the multi-${\rm D}_1$ {\it Ans\"atze} (left panels, solid lines in red and orange) and the multi-${\rm D}_2$ {\it Ans\"atze} (right panels, solid lines in blue, green and purple) are plotted for (a) low temperature ($k_{\rm B}T/\omega_c=0.01$) and (b) high temperature case ($k_{\rm B}T/\omega_c=0.2$).
	The black circles indicate the QUAPI results with $\Delta t=0.125$ and $k_{\rm max}=7$.
	The other parameters are fixed to be $\epsilon=0$, $\Delta=0.1\omega_c$ and $\alpha=0.05$.
	}
	\label{fig2}
\end{figure}%%%%%%%%%%%%

In this subsection, we focus on the sub-Ohmic regime.
Figures~\ref{fig2}(a) and \ref{fig2}(b) present results of $s=0.8$ at $k_{\rm B}T/\omega_c=0.01$ and $k_{\rm B}T/\omega_c=0.2$, respectively.
These results demonstrate that the variational approach with the multi-${\rm D}_1$ {\it Ans\"atze} provides excellent dynamics simulation over a range of temperatures and bath spectral exponents $s$.
Calculated dynamics in Fig.~\ref{fig2} shows the fast coherence decay compared to the Ohmic bath case, although the sub-Ohmic regime corresponds to the slow bath regime.
This fast coherence decay can be explained as follows.
A direct measure of the coupling strength between the system and the bath is the bath reorganization energy, $E_{r}= \int_{0}^{\infty} d\omega J(\omega)/ \omega$, which represents the dissipated environment energy after electronic excitation in accordance with vertical Franck-Condon transition in the electron transfer theory.\cite{ishizaki2010quantum}
From the definition of spectral density of Eq.~(\ref{eq6}), the value of $E_r$ in the case of $s=0.8$ is large compared with the Ohmic bath case when other parameters except the value of $s$ are fixed.
A large $E_r$ corresponds to larger fluctuations according to the fluctuation-dissipation relation,\cite{ishizaki2010quantum,kubo1985statistical} which
eradicate electronic coherence similar to what occurs in the Ohmic case.
This physical explanation is consistent with the numerical fact that the required multiplicity in both multi-${\rm D}_1$ and multi-${\rm D}_2$ {\it Ans\"atze} in the case of $s=0.8$ is not dissimilar to the Ohmic case.

\begin{figure}%%%%%%%%%%

 	 \includegraphics{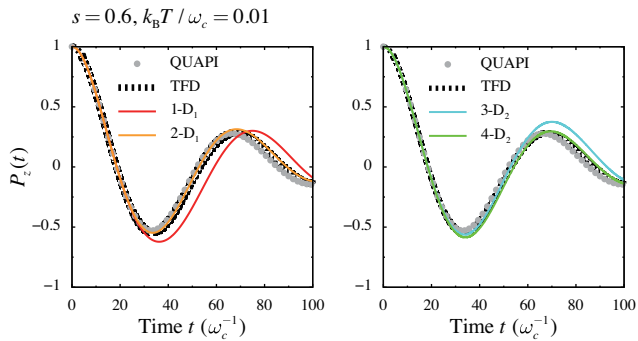}

	\caption{
	Time evolution of the population difference $P_z(t)$ with spectral density exponent $s=0.6$.
	The factorized bath initial condition is employed.
	The results calculated by the multi-${\rm D}_1$ {\it Ans\"atze} (left panels, solid lines in red and orange) and the multi-${\rm D}_2$ {\it Ans\"atze} (right panels, solid lines in blue and green) are plotted for low temperature case ($k_{\rm B}T/\omega_c=0.01$).
	The grey circles indicate the QUAPI results with $\Delta t=0.125$ and $k_{\rm max}=8$.
	The TFD approach with ${\rm D}_1^{M=4}$ {\it Ansatz} (dashed line in black) is plotted.
	The other parameters are fixed to be $\epsilon=0$, $\Delta=0.1\omega_c$ and $\alpha=0.05$.
	}
	\label{fig3}
\end{figure}%%%%%%%%%%%%

Finally, we discuss potential applicability of the multiple Davydov {\it Ans\"atze} in the deep sub-Ohmic regime ($s < 0.5$), which corresponds to situations with the ultra slow bath relaxation.
In the slow bath regime, it is known that some notable discrepancies exist
between exact results and those from the QUAPI procedure at long times, as discussed in Ref.~\onlinecite{chen2016inchworm}.
To obtain correct long time dynamics, one would need to increase the memory window size $k_{\rm max}$, making QUAPI convergence more difficult to achieve.
It has been demonstrated that the multi-${\rm D}_1$ {\it Ansatz} is consistent with results from the second-order cumulant time-nonlocal quantum master equation and the real-time path integral Monte Carlo approaches for $s=0.25$ at zero temperature, and is a reliable, efficient method for describing quantitatively accurate dynamics of the deep sub-Ohmic regime.\cite{wang2016variational}
From Figs.~\ref{fig1} and \ref{fig2}, the validity of the multiple Davydov {\it Ans\"atze} at finite temperatures also seems to be independent of the exponent $s$, but the require multiplicity may depend on it.
In order to clarify the relation between the spectral exponent and the required multiplicity, we explore a case with an exponent $s$ smaller than that in Fig.~\ref{fig2}.
As a helpful reference, we also compare our results with those from the thermal field dynamics (TFD),\cite{chen2017} similar to the approach of Borrelli and Gelin.\cite{borrelli2016}
Figure~\ref{fig3} presents result of $s=0.6$ at $k_{\rm B}T/\omega_c=0.01$.
The dynamics by ${\rm D}_1^{M=2}$ {\it Ansatz} is in perfect agreement with those of QUAPI and TFD, as shown in left panel of Fig.~\ref{fig3}.
As demonstrated in the right panel of Fig.~\ref{fig3}, it is found that the multi-${\rm D}_2$ {\it Ansatz} slightly overestimates the amplitude of population oscillation at $M=3$, and one needs to increase the multiplicity further to $M=4$ for numerical convergence.
Fortunately, no large increase in the multiplicity of the multi-${\rm D}_2$ {\it Ansatz} is needed for adequate convergence with decreasing $s$.
To explore how the required multiplicity is affected by thermal fluctuations, we have considered a case of relatively small system- bath coupling strength $\alpha$ in all numerical calculations of this study. From some previous studies of the multiple Davydov {\it Ans\"atze},\cite{wang2016variational,zhou2016fast} it has been found that the required multiplicity is reduced by increasing the system-bath coupling strength because the large bath-induced fluctuations eradicate the superposition of single Davydov states. Especially in the mod- erate to strong coupling regimes, our results suggest that both the multi-${\rm D}_1$ and the multi-${\rm D}_2$ {\it Ans\"atze} hold an advantage over the QUAPI method in the sub-Ohmic regime, where our approach is not burdened with significantly increased computational cost as the spectral density exponent $s$ is reduced.
Therefore, the multiple Davdov {\it Ans\"atze} including temperature effects is expected to be a reliable, efficient tool 
for exploring quantum dynamics in the sub-Ohmic regime at both low and high temperatures.

%%%%%%%%%%%%%%%%%%%%%%%%%%%%%%%%%%%%%%%%
\section{Concluding remarks}
In this study, we have extended the Dirac-Frenkel time-dependent variational approach with the Davydov {\it Ans\"atze} to finite temperatures by adopting the Monte Carlo importance sampling method.
To demonstrate the applicability of this approach, we have compared real-time quantum dynamics of the spin-boson model calculated by our approach with that from the numerically exact QUAPI approach.
It is shown that our variational approach with the multiple Davydov {\it Ans\"atze} is accurate over a range of temperatures and bath spectral densities.
Variational dynamics with the single Davydov {\it Ansatz} shows excellent agreement with the QUAPI results at high temperatures, while the difference between both approaches becomes significant at low temperatures.
Accuracy in dynamical calculations employing a multiple Davydov trial states is improved significantly over the single Davydov {\it Ansatz}, especially in the low temperature regime.
The reduction in the {\it Ansatz} multiplicity due to thermal fluctuations cancels out the computational cost increase due to an increased number of bath modes and extended sampling at high temperatures, and thus our variational technique with importance sampling remains efficient even at an elevated temperature.
Our results in the sub-Ohmic regime demonstrate great advantage of the variational approach with the multiple Davydov {\it Ans\"atze} because conventional perturbative approaches fail to describe strongly non-Markovian dynamics due to the long-time tail  of the time correlation function of the sub-Ohmic bath.
Our variational approach including temperate effects is expected to open up new avenues of probing quantum dynamics in the sub-Ohmic regime.

{\bf The novel advantage of the wave function propagation methods is to give access to the dynamics of all bath DOFs explicitly.\cite{schulze2015explicit,schulze2016multi,sun2010dynamics,fujihashi2016effect,zhou2015polaron} 
K{\"u}hn and coworker have investigated impacts of the quantum mechanically mixed electronic and vibrational states on electronic energy transfer dynamics in the Fenna-Matthews-Olson pigment-protein complex by tracking time evolution of bath DOFs based on the ML-MCTDH approach.\cite{schulze2015explicit,schulze2016multi}
Their calculations clearly showed the importance of vibrational motion on the local electronic ground states in the quantum mixing of electronic and vibronic excitations, which is consistent with the argument in Ref.~\onlinecite{fujihashi2015impact}. 
However, the zero temperature assumption may lead to unreliable predictions for their role at physiological temperatures because the mixed electronic and vibrational states are fragile against thermal fluctuations.\cite{fujihashi2015impact}
Our finite-temperature time-dependent variational approach with the multiple Davydov states is demonstrated to remain efficient even at an elevated temperature, and thus can explore effects of thermal fluctuations on the mixed electronic-vibrational states by tracking dynamics of vibrational DOFs.}

The time-dependent variational approach with importance sampling requires averaging over a large number of realizations at high temperatures, which may increase the computational cost in comparison with zero-temperature cases despite that the multiplicity $M$ required for convergence decreases with the increasing temperature.
Developing efficient techniques of importance sampling holds the key to improved statistics.
By employing the thermo field dynamics approach\cite{borrelli2016}, the variational approach 
with the Davydov {\it Ans\"atze} can be applied to finite temperature scenarios while avoiding the sampling procedure.
A comprehensive study along this direction is currently in progress.\cite{chen2017}

%%%%%%%%%%%%%%%%%%%%%%%%%%%%%%%%%%%%%%%%
\begin{acknowledgments}
One of us (L.W.) would like to thank Ke-Wei Sun for helpful discussion.
This work was supported by the Singapore National Research Foundation through the Competitive Research Programme (CRP) under Project No.~NRF-CRP5-2009-04 and the Singapore Ministry of Education of Education through the Academic Research Fund (Tier 2) under Project No. MOE2014-T2-1-099.
\end{acknowledgments}

%\appendix
%%%%%%%%%%%%%%%%%%%%%%%%%%%%%%
%\section{  \label{appendix-equivalence}}
%\renewcommand{\theequation}{\ref{appendix-equivalence}.\arabic{equation}}
%\setcounter{equation}{0}

\end{document}